\documentstyle[namedreferences]{spackap}

\begin{opening}
\title{Extragalactic Abundances of Hydrogen, Deuterium and Helium}
\subtitle{New Steps, Missteps and Next Steps}

\author{Craig J.  \surname{Hogan}}
\institute{Departments of Physics and Astronomy,
University of Washington, Box 351580, Seattle, WA 98195, USA}

\date{}

\end{opening}

\runningauthor{C. Hogan}
\runningtitle{Extragalactic Abundances}

\begin{document}


\begin{abstract}
Estimates of  the   deuterium abundance in quasar absorbers  are
reviewed, including  a  brief account  of  
incorrect claims published  by the author and a brief
review of the problem of hydrogen contamination. It is 
concluded that the primordial abundance may be universal with a value
$(D/H)_P\approx 10^{-4}$, within about a factor of two,
corresponding to $\Omega_B h_{0.7}^2\approx 0.02$ 
or $\eta_{10}\approx 2.7$
in the Standard Big Bang.
This agrees  
with   current limits on  primordial
helium, $Y_P\le 0.243$, which are shown to be surprisingly
insensitive to models of stellar enrichment.
It also agrees with a tabulated 
sum of the total density of baryons in observed components.
Much lower primordial deuterium 
($\approx 2\times 10^{-5}$) is also possible but
disagrees with currently estimated helium abundances;
the larger baryon density in this case fits better with current models of
the Lyman-$\alpha$ forest but requires
the bulk of the baryons to be in some currently uncounted form.
\end{abstract}


\section{Introduction}
Material in the Earth and other planets, the solar neighborhood, and
 indeed our
entire Galaxy has experienced significant chemical evolution 
that has
modified the traces of light elements from the Big Bang.
We are now beginning to sample abundances in  more distant
environments with a variety of different chemical histories.
 Some  are nearly pristine 
and serve as fossil beds preserving the original chemistry,
particularly those at
at high redshift before much of the primordial
gas first formed into galaxies and stars. 
In addition to   insights about chemical evolution
under different circumstances the new measurements
give us unique information about the  history of fine-grained
structure over
 an enormous
spacetime volume; for example they test the idea
that primordial chemistry is the same everywhere and that
the primordial gas was precisely uniform on small scales. 
New techniques also now allow  us to tabulate a more
reliable  direct estimate   of
the mean baryon density of the Universe in various forms, allowing
a further test  of Standard Big Bang Nucleosynthesis for which 
mean total baryon density  is the principal parameter. 

The theory and its concordance have been extensively 
reviewed in the literature from many points of view 
(e.g. Walker {\it et al.}   1991,
  Smith  {\it et al.} 1993,
Copi  {\it et al.} 1995,  Sarkar 1996, Fields  {\it et al.} 1996,, Hogan 1997,
Schramm 1998),
and many of the topics covered here are reviewed in more
detail elsewhere in this volume.
This review explores a selection of
interesting contradictions among
new extragalactic  datasets but also 
highlights the general concordance
 with the Standard Big Bang Model.

\section{The Cosmic Deuterium Abundance}

The measurement of deuterium abundances by analyzing Lyman series 
absorption lines  in stellar spectra
from foreground diffuse gas (Rogerson and York 1973) has yielded
 a reliable abundance for  the  Galactic 
interstellar medium, accurate to about $\pm 50\%$.
Thanks to new technology the same technique  
applied to quasar spectra now yields the abundance of
much more distant gas
(see   Tytler 1998, Vidal-Madjar 1998). 
 High resolution, high signal-to-noise spectra allow the study of 
resolved absorption
features of low optical depth which accurately count atomic column densities
and thus in principle give reliable abundances.
The subject is still very young however;
effects of  finite resolution and saturation still
 contribute a subtle source of error (see 
Levshakov et al. 1998), and the systematic errors
are not yet well calibrated since the physical model
of the absorbing material is not mature.
(It is certainly not, as assumed in the spectral fits, in discrete
isothermal slabs).  At the moment there is
an apparent polarization between ``high'' and ``low'' values
which seems likely to disappear with time.

\subsection{Evidence for a High Abundance}

An early Keck spectrum of  QSO  0014+813  allowed the first estimate
of an extragalactic deuterium abundance, yielding a remarkably
high value $D/H\approx 2\times 10^{-4}$ (Songaila {\it et al.} 1994). 
Multiple lines of the Lyman series gave a good agreement
of hydrogen and deuterium redshift, and a reliable estimate
of both column densities, so the most significant uncertainty
in this measurement was the amount of contamination 
 of the deuterium absorption feature
by interloping hydrogen.

Subsequent analysis of this same data (Rugers and Hogan 1996a) 
showed that at full resolution the Lyman-$\alpha$ feature
of deuterium was   split into two components, each of which
was too narrow for hydrogen. The authors used this fact to 
argue that the hydrogen contamination did not dominate
the  deuterium feature and therefore that the high abundance
could be trusted.
The split of the feature was traced back to an electron
excess in the raw echellogram data so was not a reduction artifact.

However, in a classic illustration of 
systematic error 
it now appears that the feature was not a real spectral
feature of the quasar.
Subsequent spectra by both the Hawaii group and by Tytler {\it et al.} (1997),
which have better signal-to-noise than the original spectrum,
do not confirm it. Apparently the $\approx 60$ electrons
contributing to the earlier signal, whatever their
origin, were neither simply noise nor photoelectrons
from the quasar light.
(Similar problems also appear to plague another candidate 
feature in the same spectrum, proposed by Rugers and Hogan 1996b).
 The deuterium feature in the real spectrum is instead smooth
and is well fit by a single thermally broadened component,
so the linewidth is consistent with significant
 hydrogen contamination. On the other hand the width is
the same as the associated hydrogen (24 km/sec) and is therefore
also  consistent
with being caused by deuterium if the broadening is mostly 
turbulent. 

Indeed, the better data allow the redshift   of the deuterium and
hydrogen features to be compared more precisely and they  
differ by 10 km/sec,
 indicating that there is  some hydrogen contaminating
the deuterium (Tytler {\it et al.} 1997). 
However the total column density of   contaminating hydrogen
required to move the centroid 
is   small, so the known presence of some contamination
 does not imply that the best estimate of the abundance 
changes appreciably from that originally given 
by Songaila {\it et al.}

Because of the good constraints on both hydrogen and deuterium
column densities, and in spite of  confusing claims made by
this author, the new data on  the Q0014+813 absorber   still  
displays good evidence of a high primordial abundance, about
as convincing as the  original  claim   by Songaila {\it et al.} (1994).
In another quasar   (BR 1202-0725) a ``detection or upper limit'' 
at $D/H= 1.5\times 10^{-4}$ was found by
 Wampler {\it et al.} (1996).
A high abundance ($2\times 10^{-4}$) may also be detected
in Q0402-388, although it  is 
required only if OI/HI is assumed to be constant in the fitted
components (Carswell {\it et al.} 1996).
The same  high abundance ($2\times 10^{-4}$)
was also  found by Webb {\it et al.} (1997)
in Q1718+4807, although this is also not yet conclusive;
the present analysis    relies on a SiIII line to fix
the redshift of the hydrogen,   the D column is
 based only on a Lyman-$\alpha$ fit 
and the H column on a
 low resolution spectrum of the Lyman limit. 
The agreement between these estimates is at least suggestive
 of a high universal abundance,
although none of the evidence is yet conclusive.

\subsection{Evidence for a Low Abundance}
 
Burles and Tytler (1996) and Tytler {\it et al.} (1996) have presented
evidence for a low $D/H$ in two quasar absorbers.  Of these the stronger
case at present is in Q1937-1009 since high quality data are 
available up to the Lyman limit. The estimated abundance is
$2.3\pm 0.3\times 10^{-5}$, nearly an order of magnitude less than
the high values discussed above.

 Unfortunately the total column in this
case is high so the HI absorption is optically thick even past
the Lyman limit and the column density must be estimated 
from saturated features.  This has led to a debate in the literature
on the allowed range for the HI column and   for $D/H$
(e.g. Songaila {\it et al.} 1997,
Burles and Tytler 1997, Songaila 1997), which is likely to end up
somewhere in the middle: although the total error in the
abundance is probably larger than originally quoted by
 Tytler {\it et al.} (1996),
the HI column is probably well enough constrained to exclude
very  high values like those in Q0014+813.

Therefore the dispersion in abundance between the best high and
low estimates 
appears on the surface to be real.
What is going on? There have been many suggestions
(e.g., Jedamzik and Fuller 1996). Perhaps
the primordial abundance is not uniform; perhaps the low-D
systems have experienced stellar destruction of their 
deuterium; perhaps the high-D systems have found
some exotic source of nonprimordial deuterium.
 The most prosaic explanation however is
that the high-D features are all dominated by contaminating
hydrogen.

\subsection{Statistical Approaches to the Contamination Problem}

The latter possibility deserves serious attention since the effect
is known to be there and known to bias abundance estimates upwards.
However, a quantitative estimate of the effect shows that it is 
unlikely to be important most of the time.
The mean number of hydrogen lines in a velocity interval
$\delta v$  per $\ln(N[HI])$
interval can be estimated from the line counts of Kim {\it et al.}
(1997) to
be about
$$
\delta P(N[HI])\approx 5\times 10^{-3}\left({\delta v\over 10{\rm  km\  sec^{-1}}}\right)
\left({N[HI]\over 10^{13}{\rm cm^{-2}}}\right)^{-0.4}.
$$
If the contaminating hydrogen lines are distributed at random,
we can use  Poisson statistics  to estimate the
probability of  contamination.
For example, in  the case of
Q0014+813 the column density of the DI feature is about
$10^{13.2}{\rm cm^{-2}}$; the probability of an HI line
close to this column density    appearing in  the right redshift range
to mimic deuterium (within an interval of about 
$\delta v\approx 20{\rm  km\ \sec^{-1}}$) is only about $P\approx 1\%$. 

Of course, smaller amounts of contamination are more likely.
They  tend to bias the deuterium abundance estimates upwards and create
a nongaussian (power-law) error distribution
allowing low $D/H$ with nonnegligible probability.
 However the magnitude
 of the bias is still small in this range of column density; for example,
the chance of a $\approx 10\%$ contamination (for which there is indeed 
some evidence in Q0014+813 in the line profiles) is 
greater than the probability of 100\% 
contamination by a factor of about $10^{0.4}$; 
but this is  still only a few percent.

Furthermore this calculation
 does not yet allow for the additional coincidence required
in the Doppler parameters. Real deuterium
cannot be wider than its corresponding hydrogen. Contaminating hydrogen  
has a linewidth drawn at random from the parent population, which 
 is not compatible with a deuterium identification 
if it happens to exceed 
 the width of the corresponding hydrogen feature.
In the case of Q0014+813, since the 
D feature linewidth of 24 ${\rm km\  sec^{-1}}$
is typical of HI Lyman-$\alpha$ forest 
lines  we should multiply the probabilities
by about half to account for this effect. This
factor is   smaller in
 situations where  the   features are unusually narrow.

It is important to verify  that the lines are
uncorrelated as we have assumed.
Although there are disagreements over the amplitude
of line correlations the best data on the 
Lyman-$\alpha$ forest shows correlations  smaller than unity 
at velocity separations of $\simeq 100 {\rm km \ sec^{-1}}$;
for example
 in the line lists of Kim {\it et al.} (1997)
the amplitude is less than 10\% for all lines. We have checked 
this also specifically for correlations  
with high HI column density features similar to those studied
in the D candidates; although the sample is smaller 
and the result noisier, the amplitude of the correlation  is
still   less than  a few tenths.  In this
study we used  one-sided correlations (counting
line companions only to the red side of each Lyman-$\alpha$ line) in 
order to exclude first-order effects of deuterium contamination
in the hydrogen sample. (Similar conclusions were recently 
reached by Songaila 1997). 

Thus even without a thorough
physical understanding of the absorbing material, 
these empirical studies indicate  that correlations among the lines are too
weak to alter  significantly   the   simple 
estimates made on the basis
of Poisson statistics.  Therefore the best guess is
that the deuterium abundance in   cases where it appears to
be high really is high, at least where the DI
column is greater than about $10^{13}{\rm cm^{-2}}$
(which is in any case required for a reasonably precise
column estimate at realistic signal-to-noise).
Noting that the full error in the fitting technique
has not yet been calibrated on realistic physical
models of the clouds, it still seems reasonable to 
guess that the current data are consistent with a universal primordial
abundance  with a factor of two of $10^{-4}$. 
A much more convincing measurement will be possible
with a  few more good targets.

\subsection{Next Steps}

The contamination problem is less for lower redshift, where
the Lyman-$\alpha$ forest thins out. Although low redshift
quasar spectroscopy at high resolution 
is costly as it  requires a large
investment of time with the Hubble Space Telescope, the new STIS
two-dimensional spectrograph can observe the entire Lyman series at
the same time; the greater efficiency will give more
solid results on cases such as Q1718+4807 which are already
known to be interesting (e.g. Webb {\it et al.} 1997) .  

Contamination can also be reduced by studying high column
density systems. Especially interesting are damped
HI absorbers where reliable columns can be obtained for
both HI and DI (Jenkins 1996), although targets are also
rarer and 
 tend to be in evolved galaxies. The best target of
this type so far identified is Q2206-199,
which has a low metal abundance and a very low
velocity dispersion (Pettini and Hunstead 1990).

In the long term this problem will be solved by a larger
sample of absorbers in bright quasars observed from the 
  ground. Although the current number of
suitable target quasars is now quite small, the target sample in
coming years will 
grow by about two orders of magnitude  as a result
of the Sloan Digital Sky Survey,  so progress in this field
will  be limited by observing time rather than by
the availability of targets.

\section{The Cosmic Helium Abundance}

\subsection{A Bayesian Approach to Helium Enrichment}
Helium abundances in extragalactic HII regions have
for many years been the principal observational constraint
on Big Bang Nucleosynthesis.
Flourescent nebular emission lines of hydrogen and helium
reveal quite precisely the number of electrons recombining
into each species and thereby the abundances of each 
(e.g. Peimbert and Torres-Peimbert 1976, 
Pagel et al. 1992, Skillman and Kennicutt 1993).
 The 
techniques and estimates of  systematic
errors are discussed by H{\o}g et al. (1998), Skillman et al. (1998),
and Steigman et al.(1998). Here I make one simple point:
the dominant source of systematic error
 in the primordial abundance $Y_P$, especially 
in the upper limit, probably lies not in the model used to extrapolate
to zero metallicity but in the physical models used to estimate
the present-day abundances $Y $ from observations of nebulae.

Even though the bulk of  the helium of the Universe originates in the Big Bang,
the additional helium enrichment by  stars cannot be ignored 
in estimating the primordial abundance 
from observations of present-day helium.   
The most widely used approach  to estimate the nonprimordial enrichment
is to correlate $Y$ with metallicity $Z$.
However, one of the limitations of this method is the need to
 assume
 a linear relation between $Y$ and $Z$, 
which is not well motivated. 
Moreover most of the information on the primordial abundance is contained
in the lowest metallicity points, where the correlation is 
not very reliably established; this information is not 
being efficiently used in regression fits dominated
by highly enriched regions.
Another approach has been to simply take 
the lowest, best measured points and use them as estimates of
(or at least limits on) the primordial abundance. Clearly however
some correction has to be made for the obvious 
Malmquist-like bias introduced,
and some statistical
 way needs to be found to combine more than one region.

Hogan, Olive and Scully (1997)  introduced
 a new statistical method to estimate the 
primordial helium abundance $Y_p$ directly from helium
observations, 
without using metal abundances.    They
constructed a likelihood function using a Bayesian prior, 
encapsulating the key assumption  that the true
helium abundance must always exceed the primordial value,
by an amount which may be as large some maximum enrichment $w$.
They computed the likelihood as a function of the two parameters 
$Y_p$ and $w$ using  samples of   measurements compiled from
the literature.

\begin{figure}
\vspace{2in}
\caption{Likelihood   function  
showing $1\sigma$, $2\sigma$ and $3\sigma$ contours in the 
$(Y_p,w)$ plane, from Hogan, Olive and Scully (1997).
  The +'s indicate
the peaks of the likelihood functions. The two left  panels 
show results using a top-hat prior and 
two different subsamples of 11 and
  32 lowest metal points. The two right panels show the results
for positive and negative bias priors, for the 32 point
sample. In all cases the  conservative 2$\sigma$ limit
occurs at $w=0$ and yields a limit $Y_P\le 0.243$.}
\end{figure}

Some results from published samples are shown in
figure 1.  Estimates of 
$Y_p$ vary between 0.221  and 0.236,
depending on the specific subsample and  
prior adopted,    consistent with previous
estimates  using different techniques.
Evidence for stellar enrichment ($w\ne  0$) appears 
even in the lowest metallicity subsamples, but in all samples
the most conservative upper bound on $Y_p$ occurs
for $w=0$, 
yielding a nearly model-independent bound
 $Y_p < 0.243$ at 95\% confidence. 
 The main uncertainty in the $Y_p$ bound
is not
the model of stellar enrichment but possible common systematic
biases in the estimate of $Y$ in each individual HII region.

\subsection{Helium at High Redshift}

 In highly ionized environments  
   singly
ionized helium (HeII) is typically orders of magnitude
 more common than HI, making it
the most cosmically  abundant absorber, detectable even
in the most rarefied regions between the  Lyman-$\alpha$ forest  clouds.
  Lyman-$\alpha$ absorption by  HeII nearly continuously
fills redshift space at optical depth of the order of unity
as recently verified   in at least three quasars
(Jakobsen {\it et al.} 1994, Davidsen {\it et al.} 1996, 
Hogan {\it et al.} 1997, Reimers {\it et al.} 1997).

The need to model ionization
states precludes a truly precise measurement of helium abundance.
An absolute helium abundance can be estimated from
HI and HeII Lyman-$\alpha$ absorption 
 if:
(1)  the helium abundance is uniform; (2)  
HeII and HI are in ionization 
equilibrium dominated by photoionization; (3)
we know the shape of the ionizing spectrum; 
(4) absorption is unsaturated, so column densities
of absorbing species can measured.
The total redshift integrated columns of  HeII and HI are then
both proportional to the same 
line integral $\int d\ell n_e^2$ with coefficients depending
on the abundance and the ionizing spectrum.

These conditions are certainly never met in detail,
but in some situations data can still be used to set  constraints on
the  abundance and its variations.
 In some regions, the HeII fraction may be close to unity
(Reimers {\it et al.} 1997); in others, the strength and/or 
spectrum of the  ionizing field may be known, for example
for gas 
close to the quasar (Hogan  {\it et al.} 1997); and
in some circumstances the spectrum   may be   approximately
constant over a large pathlength along the line of sight,
allowing the universality of the abundance to be 
constrained even if its absolute value is uncertain.

The new  results are however most  important
not as abundance determinations but
 as probes    of structure
formation and ionization history.
If we assume that the helium abundance
is close to the Big Bang prediction, 
  helium observations
put stringent constraints on the density of diffuse gas
at high redshift which limit the range of possible
baryonic histories. For example, it is not possible to place 
the bulk of the baryons in
a diffuse medium  at $z=3$ without exceeding helium absorption 
limits unless the gas is hot enough to be thermally 
ionized; on the other hand a substantial fraction of 
the baryons appear to be necessary in gaseous form
(in clouds) to account for the Lyman-$\alpha$ forest
(Rauch {\it et al.} 1997, Weinberg {\it et al.} 1997, 
Zhang {\it et al.} 1997).

\section{The Cosmic Baryon Abundance and the
Concordance of    Standard Big Bang Nucleosynthesis}

In a recent survey, Fukugita, Hogan and Peebles
(1997) estimate the mean density of the Universe 
observed in various components of baryons, 
summarized in Table 1.
Not included   in the table are two ``uncounted'' components,
MACHOs and hot  plasma ($T\approx 2\times 10^6$K),
which are known to exist locally. For both of these almost no
meaningful upper or lower bounds can be set on the global density.
In the case of MACHOs, a new   population
of dark compact objects, probably baryonic,
has been detected by microlensing in the direction of the LMC;
depending on assumptions used to extrapolate,
the global density of this population could either be negligible
or could dominate all other forms of baryons combined. Similarly,
a thermal background from hot gas is detected, but could
either be from a globally insignificant portion of the Galactic 
corona or from a globally distributed plasma containing
the bulk of the baryons. These two components are therefore
possible repositories of additional baryons.

The point here of course is to compare the observed
number of baryons with the number expected on 
the basis of light element abundances.
 For the nucleosynthesis entries we adopt upper
and lower limits for the primordial deuterium
$2\times 10^{-5}\le (D/H)_P\le 2\times 10^{-4}$.
For primordial helium we adopt a central value of
$Y_P= 0.23$ and a 2$\sigma$ limit as
described above, $Y_P\le 0.243$. Note that this
limit is not compatible with the low deuterium
values.  The lithium
abundance allowing for some depletion
 is taken to be $Li/H\le 4\times 10^{-10}$
from Galactic stars (see Cayrel 1998);  
even though it does not offer a principal constraint 
on baryon density in the Standard Model, it  
 is important as a constraint  on departures
from the Standard Model, such as small scale inhomogeneities
(e.g. Kurki-Suonio {\it et al.} 1997).
\begin{table}
 
\caption{The Baryon Budget (Fukugita {\it et al.} 1997)}
\begin{tabular}{llllll}\hline
 {} &  {Component} & {Optimum} & 
 {Maximum} & {Minimum} 
&  {Grade}\\ 
\hline \\
 
  & observed at $z\approx 0$: &   \hfil   &   \hfil  &  \hfil    &  \hfil  \\
1 & stars in spheroids &0.0026\hfil$h_{0.7}^{-1}$\hfil&  0.0043 $h_{0.7}^{-1}$
      &   0.0016  $h_{0.7}^{-1}$  &A  \\
2 & stars in disks &0.00086\hfil$h_{0.7}^{-1}$\hfil&  0.00129 $h_{0.7}^{-1}$ 
&  0.00051  $h_{0.7}^{-1}$  &A--  \\
3 & stars in irregulars &0.000069\hfil$h_{0.7}^{-1}$\hfil& 
0.000116 $h_{0.7}^{-1}$
& 0.000033  $h_{0.7}^{-1}$  &B  \\
4 & neutral atomic gas &0.00033\hfil$h_{0.7}^{-1}$\hfil&  0.00041 $h_{0.7}^{-1}$ 
&  0.00025 $h_{0.7}^{-1}$ &A  \\
5 & molecular gas &0.00022\hfil$h_{0.7}^{-1}$\hfil&  0.00029 $h_{0.7}^{-1}$ 
&  0.00014 $h_{0.7}^{-1}$ &A--   \\
6 & plasma in clusters &0.0026\hfil$h_{0.7}^{-1.5}$\hfil&  0.0044
$h_{0.7}^{-1.5}$  &0.0014\hfil$h_{0.7}^{-1.5}$\hfil&A   \\
$7$ & plasma in groups &  0.014\hfil$h_{0.7}^{-1}$\hfil&  0.03
 $h_{0.7}^{-1}$
&  0.007 $h_{0.7}^{-1}$ &B \\
\cr \cline{1-6}\cr
 & sum (at $h=0.7)$&  0.02    &  0.04    &  0.01   &     \\
\hline\\
& observed at $z\approx 3$: 
 &   \hfil   &   \hfil  &  \hfil    &  \hfil  \\
10 &   damped absorbers & 0.001-- 0.002 $h_{0.7}^{-1}$   &  0.0027  $h_{0.7}^{-1}$
      &   0.007  $h_{0.7}^{-1}$  &   A--  \\
11 &  Lyman-$\alpha$ forest clouds &     0.038 $h_{0.7}^{-2}$  &   
&  0.025  $h_{0.7}^{-2}$  &    B  \\
12 & intercloud gas (HeII) &  &  0.01 $h_{0.7}^{-1.5}$
& 0.0001  $h_{0.7}^{-1}$  &B  \\
\hline\\
 & nucleosynthesis:&   &   &   &  \\
9a & deuterium &   & 0.054 $h_{0.7}^{-2}$ & 0.013 $h_{0.7}^{-2}$ & A \\
9b & helium& 0.010 $h_{0.7}^{-2}$ & 0.027 $h_{0.7}^{-2}$ &   & A-- \\
9c & lithium &   & 0.06 $h_{0.7}^{-2}$ & 0.007  $h_{0.7}^{-2}$ & B \\
\hline\\
\end{tabular}
\end{table}

It is clearly instructive to contemplate these numbers at
length.   
Most of the baryons today are still in the 
form of ionized gas, which contribute a mean density uncertain
by a factor of about four (due to uncertainties in
extrapolating from observed x-ray emission). For the best-guess plasma
density,  
 stars are a relatively minor component--- all stars and their
remnants  comprise only about $17\%$ of the baryons, while
populations contributing most of the blue light comprise
less than 5\%. The formation of galaxies and of stars within
them  appears to be a globally   inefficient process---
an effect not fully understood in models of
galaxy formation.
The sum over the budget, expressed as a fraction
of the critical Einstein-de~Sitter density, is in the range 
$0.01\le \Omega _B\le 0.04$,
with a best guess $\Omega _B\sim 0.02$ (at Hubble parameter
70~km~s$^{-1}$~Mpc$^{-1}$). This is close to the prediction from
the Standard Big Bang
for moderately high $D/H\approx 10^{-4}$.
If the deuterium abundance is high, this suggests we may be 
close to a complete survey of the major states of the baryons
as well as a concordance with nucleosynthesis.
On the other hand if $D$ is low (and $Y_P$ has been
underestimated) the baryon budget is likely to be dominated by
currently uncounted components. Although galaxy
formation models prefer the higher baryon density
(e.g. Rauch {\it et al.} 1997, Zhang {\it et al.} 1997) they do not
yet predict   which of
the hidden forms dominates today (see e.g. Fields and Schramm 1998)--- 
very diffuse gas or very 
compact cold bodies.  

\acknowledgements
I am grateful  
 to my  collaborators, especially S. F. Anderson, 
M. Fukugita, K. A. Olive, M. H. Rugers, P. J. E. Peebles,
and S. T. Scully, for many insights, and A. Songaila and L. L.
Cowie for generous sharing of their superb  data.
 This work was supported by NASA and
NSF at the University of Washington.

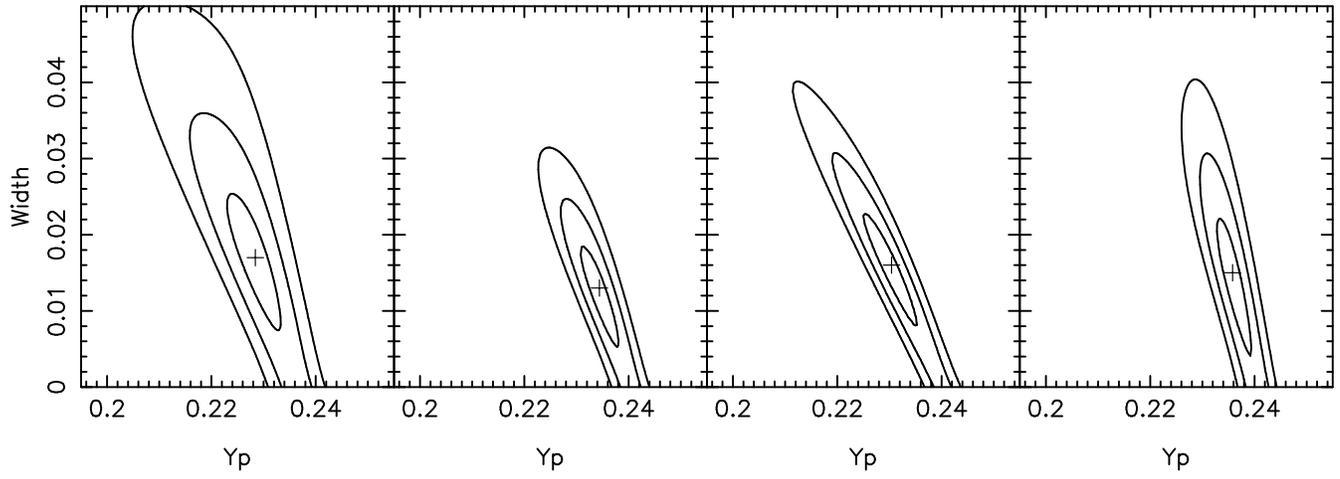
\end{document}